\documentclass[12pt,preprint]{aastex}
\usepackage{bbm}
\usepackage{mathrsfs}
\usepackage{amssymb,amsmath,float}

\newcommand\al{\alpha}

\newcommand\de{\delta}

\newcommand\rh{\rho}

\newcommand\ta{\tau}

\newcommand\om{\omega}

\newcommand\De{\Delta}

\newcommand\Ph{\Phi}

\newcommand\<{\langle}
\renewcommand\>{\rangle}

\newcommand\ie{\emph{i.e.}}
\newcommand\eg{\emph{e.g.}}

\newcommand\beq{\begin{equation}}
\newcommand\eeq{\end{equation}}
\newcommand\bea{\begin{eqnarray}}
\newcommand\eea{\end{eqnarray}}
\newcommand\bal{\begin{align}}
\newcommand\eal{\end{align}}

\newcommand\fr{\frac}

\newcommand\half{{\textstyle \frac{1}{2}}}
\newcommand\ap{\approx}

\renewcommand\bal{\mbox{\boldmath$\alpha$}}

\newcommand\cT{\mathcal{T}}

\begin{document}

\title{A method to improve the sensitivity of radio telescopes}

\author{Richard Lieu$^1$,  T.W.B. Kibble$^2$, and Lingze Duan$^1$}

\affil{$^1$Department of Physics, University of Alabama,
Huntsville, AL 35899\\
$^2$Blackett Laboratory, Imperial College, London SW7~2AZ, U.K.\\}



\begin{abstract}
As an extension of the ideas of Hanbury-Brown and Twiss, a method is proposed to eliminate the phase noise of white chaotic light in the regime where it is dominant, and to measure the much smaller Poisson fluctuations from which the incoming flux can be reconstructed.  The best effect is achieved when the timing resolution is finer than the inverse bandwidth of the spectral filter.  There may be applications to radio astronomy at the phase noise dominated frequencies of $1 - 10$~GHz, in terms of potentially increasing the sensitivity of telescopes by an order of magnitude.
\end{abstract}

\section{Introduction and background}

In astronomical observations, depending upon the radiation bandpass of interest usually one of the two natural components of flux uncertainties, {\it viz.} phase and Poisson fluctuations, plays the principal role of masking the genuine incoming signal of celestial sources.  Specifically, while at radio frequencies of photon abundance classical phase noise dominates any estimate of the necessary exposure time (\eg~\cite{bur10}), at higher frequencies it is the Poisson counting statistics of individual photons that matter (\eg~\cite{bir06}).  The intention of this paper is to propose a method of eliminating the phase noise fluctuations in the wavelength region where they are dominant.  While of little significance to efforts in improving sensitivities in the optical and X-ray telescopes, there may be applications to radio astronomy.

To setup the background, we consider a unidirectional beam of cross-sectional area $A$ propagating in the $z$ direction, and for simplicity assume that there is only one polarization mode, e.g., plane polarization in the $x$ direction.

In a quantum treatment, the positive frequency part of the electric field operator is given (in units with $c=\hbar=1$) by
 \beq \hat E^{(+)}(t,z) = \fr{i}{\sqrt{4\pi A}}\int d\om \sqrt{\om} \hat a(\om) e^{-i\om(t-z)}, \eeq
where $\hat a(\om)$ and $\hat a^\dag (\om)$ are respectively the annihilation and creation operators of the radiation mode of frequency $\om$;
while the negative frequency part is the complex conjugate,
 \beq \hat E^{(-)}(t,z) = \hat E^{(+)}(t,z)^\dag. \eeq
The cycle-averaged intensity operator $\hat I(t,z)$, which represents the energy flux density (the magnitude of the Poynting vector) averaged over a cycle and integrated over the area $A$, is given by
 \beq \hat I(t,z) = 2A\hat E^{(-)}(t,z)\hat E^{(+)}(t,z). \eeq

In the case of a stationary light beam, there is no correlation between different frequencies, and we may write
 \beq \<\hat a^\dag(\om)\hat a(\om')\> = n(\om)\de(\om-\om'), \label{aadagav}\eeq
where the angle brackets denote the ensemble averaged expectation value.  Thus
 \beq \bar I \equiv \<\hat I(t,z)\> = \int \fr{d\om}{2\pi}\,\om n(\om), \eeq
where $n(\om)$ is the photon occupation number of mode $\om$.
For example, if we have a Gaussian wave form with central frequency $\om_0$ and reciprocal bandwidth $\ta$,~{\it viz.}
 \beq n(\om) = \sqrt{2\pi} n_0 e^{-(\om-\om_0)^2\ta^2/2}, \label{gauss}\eeq
where
 \beq \tau = \fr{1}{\de\om}= \fr{1}{2\pi \de\nu}, \label{band} \eeq
then
 \beq \bar I = \fr{\om_0 n_0}{\ta}, \label{Iav}\eeq
so that the mean rate of arrival of photons is $n_0/\ta$.  Such a spectrum typically arises from broad band emission transmitted through a narrow filter.

\section{Origin of the two components of radio noise}

It is useful to begin with a recapitulation of the standard derivation of intensity fluctuation.  A real measurement will always occupy a finite length of time, so let us consider the intensity $I$ averaged over some time interval $T \gg 1/\om_0$
 \beq \hat I_T=\fr{1}{T}\int_0^T \hat I(t). \label{IT} \eeq
Obviously,
 \beq \<\hat I_T\> = \<\hat I\>=\bar I. \eeq

To find the variance, we consider
 \bea \<\hat I_T^2\> \!\!\!&=&\!\!\! \fr{4A^2}{T^2}\int_0^T dt_1dt_2
 \<\hat E^{(-)}(t)\hat E^{(+)}(t)\hat E^{(-)}(0)\hat E^{(+)}(0)\> \notag\\
 \!\!\!&=&\!\!\! \fr{1}{(2\pi)^2T^2}\int_0^T dt_1dt_2
 \int d\om_1d\om'_1d\om_2d\om'_2\sqrt{\om_1\om'_1\om_2\om'_2} \notag\\
 &&\ \ e^{i(\om_1-\om'_1)t_1+i(\om_2-\om'_2)t_2}
 \<\hat a^\dag(\om_1)\hat a(\om'_1)\hat a^\dag(\om_2)\hat a(\om'_2)\>. \label{expIsq}\eea
Now in the case of a beam with Gaussian statistics, the four-point function can be written as a sum of products of two-point functions:
 \bea &&\<\hat a^\dag(\om_1)\hat a(\om'_1)\hat a^\dag(\om_2)\hat a(\om'_2)\> \notag\\
 &&\ \ = \<\hat a^\dag(\om_1)\hat a(\om'_1)\>\<\hat a^\dag(\om_2)\hat a(\om'_2)\>
+\<\hat a^\dag(\om_1)\hat a(\om'_2)\>\<\hat a(\om'_1)\hat a^\dag(\om_2)\>\eea
Substituting the first term here into (\ref{expIsq}) clearly reproduces $\bar I^2$.  The variance is therefore given by the second term.  Using (\ref{aadagav}), together with the commutator
 \beq [\hat a(\om'),\hat a^\dag(\om)] = \de(\om'-\om), \eeq
we find
 \beq \De I_T^2 = \<\hat I_T^2\> - \<\hat I\>^2
 = \fr{1}{(2\pi)^2T^2}\int_0^T dt_1dt_2 \int d\om_1d\om_2\,\om_1\om_2 n(\om_1)[ n(\om_2)+1]
 e^{i(\om_1-\om_2)(t_1-t_2)}. \label{DeltaI_T1}\eeq
In practice we always have $\om_0 \gg 1/\ta$, and under those circumstances the factors of $\om_j$ can be replaced to good accuracy with $\om_0$.  Where the frequency profile is Gaussian, eq.~(\ref{gauss}), this gives
 \beq \De I_T^2 = \fr{\om_0^2}{\ta T}\left[n_0^2 F\left(\fr{T}{\ta}\right)+n_0\right],
 \label{DeltaI_T} \eeq
where
 \beq F(x) = \sqrt{\pi}\,\text{erf}(x) - (1-e^{-x^2})/x. \label{F}\eeq
We note the limiting forms
 \bea F(x) \!\!\!&\ap&\!\!\! \sqrt{\pi} - 1/x, \qquad x\gg 1, \notag\\
 F(x) \!\!\!&\ap&\!\!\! x - x^3/3,\ \,\qquad x\ll 1. \eea
The last term of (\ref{DeltaI_T}) is the shot noise term.  Equivalently,
 \beq \left(\fr{\De I_T}{\bar I}\right)^2 = \fr{\ta}{T}\left[F\left(\fr{T}{\ta}\right)+\fr{1}{n_0}\right]. \eeq

For radio observations where $n_0 \gg 1$ by virtue of the high system temperature (\ie~large background flux), the first term is dominant unless $T/\ta$ is extremely small.  For $T\gg\ta$, one has
 \beq \left(\fr{\De I_T}{\bar I}\right)^2 \ap \fr{\sqrt{\pi}\ta}{T}
 =\fr{1}{2\sqrt{\pi}T\de\nu}, \label{radio} \eeq
in accordance with the radiometer equation (\cite{chr85,bur10})

Note that the results in this section are valid for the intensity behavior of radio waves irrespective of their origin, which can be celestial, atmospheric, scattered ground radiation, or thermal noise of the receiver's preamplifiers.

\section{Phase noise cancellation by a beam splitter}

The fluctuations described in the previous section can accurately be removed by subtracting the intensity $I'$ of a compensating beam from the primary beam, as we demonstrate below.  We first assume that both beams have equal\footnote{The assumption of a 50:50 beam spliter is actually inessential to the arguments and conclusion of this paper.} mean intensity, \ie~\beq \bar I^{\rm t} = \bar I^{\rm r} = \half\bar I, \eeq because the incident beam is divided 50:50 at a beam splitter and the intensities of the two emerging beams are measured {\it without} combining them, see Figure 1. In what follows it is further assumed that the pathlengths traversed before measurement are equal, and the thickness of the beam splitter is small w.r.t.\ the radiation wavelength (an achievable criterion at radio frequencies).

\begin{figure}
\begin{center}
\includegraphics[angle=0,width=6in]{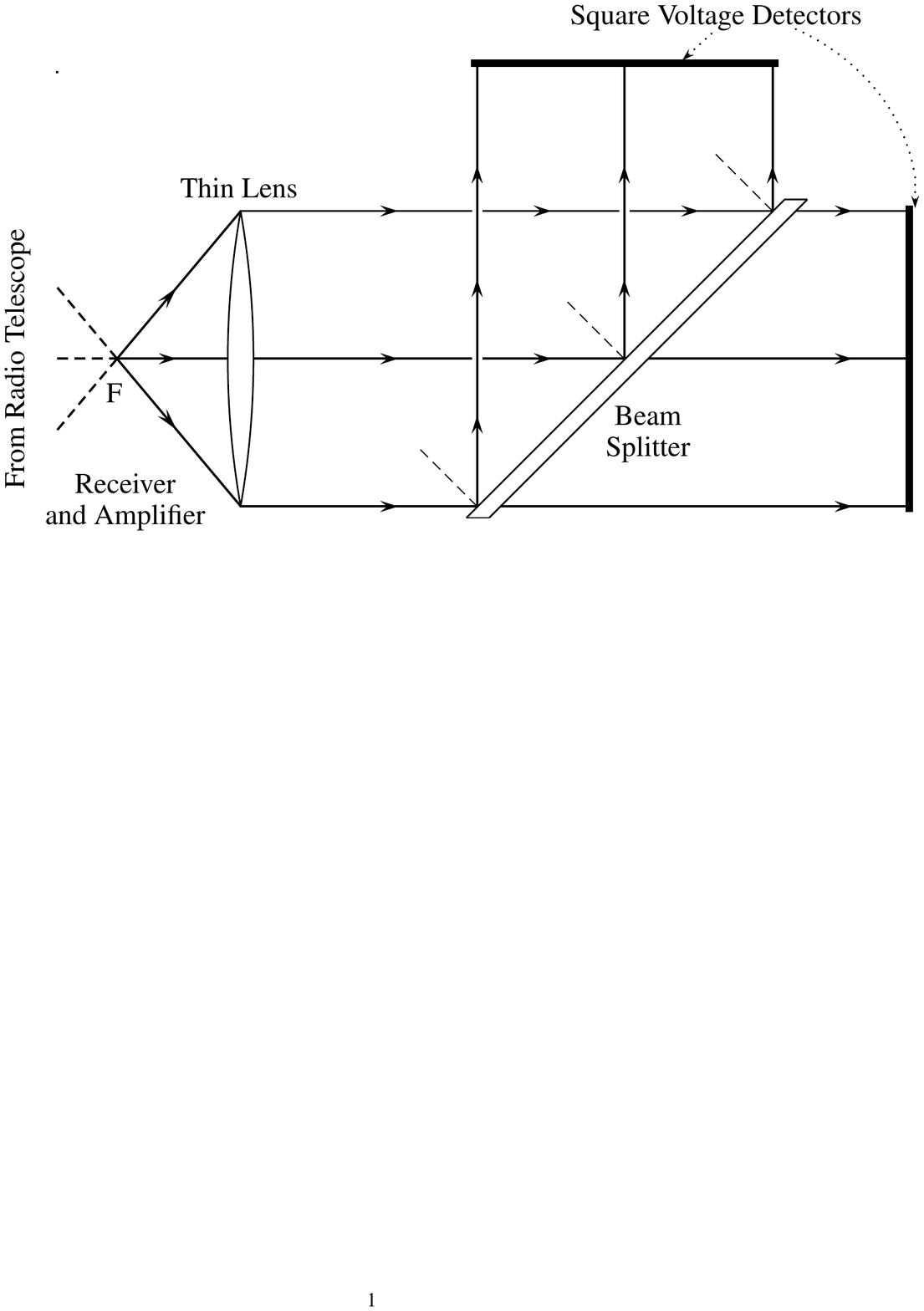}
\vspace{-7cm}
\end{center}
\caption{The radio signal from a celestial source, after it is focused by the telescope and amplified, is passed through a `Hanbury-Brown Twiss' thin beam splitter {\it without} recombining.  The intensity time profiles of the two beams are then measured after they traversed equal path lengths.}
\end{figure}

The electric fields in the two beams can be expressed in terms of creation and annihilation operators that satisfy the same commutation relations as before:
 \beq [\hat a^{\rm t}(\om),\hat a^{{\rm t}\dag}(\om')]
 = [\hat a^{\rm r}(\om),\hat a^{{\rm r}\dag}(\om')] = \de(\om-\om'), \eeq
but of course
 \beq [\hat a^{\rm t}, \hat a^{{\rm r}\dag}]=0. \label{trcomm} \eeq
The field in the reflected beam may suffer a phase change relative to the incident one, but that makes no difference because the operators appear in pairs and the overall phase cancels out.  The expectation values of products are given by expressions of the same form (\ref{aadagav}), but with
 \beq n^{\rm t}(\om)=n^{\rm r}(\om)=\half n(\om). \eeq

However, we also need to evaluate such mixed expressions as $\<\hat a^{{\rm t}\dag}(\om)\hat a^{\rm r}(\om')\>$.  Now, in the classical case, the beam can be represented as a statistical mixture, in which the field $\Ph_j$ appears with probability $p_j$.  (Of course, the labelling index here may actually be continuous rather than discrete, but it is sufficient to consider the discrete case.)  The corresponding quantum state can be taken to be a superposition of coherent states, of the form
 \beq \rh = \sum_j p_j |\al_j\>\<\al_j|, \eeq
where $|\al_j\>$ is an eigenstate of the annihilation operators, with eigenvalue defined by $\Ph_j$.

What then happens when the light hits the beam splitter?  Classically, for an input field $\Ph_j$, we end up with a transmitted field $\Ph_j^{\rm t} = \Ph_j/\sqrt 2$ and a reflected field $\Ph_j^{\rm r}$ of the same magnitude and possibly altered phase.  However, since the overall phase will always cancel out, we can ignore it and take $\Ph_j^{\rm r}=\Ph_j^{\rm t}$.  Quantum mechanically, in place of the coherent state $|\al_j\>$ of the incident beam we have a coherent state $|\al^{\rm t}_j,\al^{\rm r}_j\>$, an eigenstate of both sets of annihilation operators, with eigenvalues defined by $\Ph_j^{\rm t}$ and $\Ph_j^{\rm r}$.  When we take the expectation value of the product $a^{{\rm t}\dag}(\om)\hat a^{\rm r}(\om')$, we can then replace each of the operators by the corresponding eigenvalues, so the result is essentially the same as for pairs of operators from the same beam.  It follows therefore that as in (\ref{aadagav})
 \beq \<\hat a^{{\rm t}\dag}(\om)\hat a^{\rm r}(\om')\> = \half n(\om)\de(\om-\om'), \eeq
ignoring any overall phase which would cancel out.

For measurements of intensity averages over an interval $T$ it is clear that their variances $(\De I_T^{{\rm t}})^2$ and $(\De I_T^{{\rm r}})^2$ are each given by exactly the same expressions as (\ref{DeltaI_T}) above, but with $n_0$ replaced by $n_0/2$.  Moreover, for the cross correlation, the term quadratic in $n_0$ is the same, but there is no linear term (\ie~no shot noise, or the last term of (\ref{DeltaI_T}))  because of (\ref{trcomm}), so
 \beq \<I_T^{\rm t} I_T^{\rm r}\> -\bar I^{\rm t}\bar I^{\rm r} =
 \fr{\om_0^2 n_0^2}{4T\ta}F\left(\fr{T}{\ta}\right). \label{crosscorr} \eeq
The absence of shot noise from (\ref{crosscorr}) is the basis of the Hanbury-Brown and Twiss effect (\cite{han57}), {\it viz.} the beam splitter cannot divide photons and so each is either reflected or transmitted by random 50:50 chance.

But suppose, instead of pursuing intensity interferometry, we form the difference signal $I_T^{\rm d} = I_T^{\rm t}-I_T^{\rm r}$. Evidently its mean is zero:
 \beq \<I_T^{\rm d}\> = 0. \eeq
The variance is given by
 \beq (\De I_T^{\rm d})^2 = \< (I_T^{\rm t}-I_T^{\rm r})^2\>
 =(\De I_T^{{\rm t}})^2+(\De I_T^{{\rm r}})^2-2[\< I_T^{\rm t}I_T^{\rm r}\>
 -\bar I^{\rm t}\bar I^{\rm r}]. \eeq
It is clear that the quadratic terms in $n_0$ here will cancel exactly, leaving only the linear terms, {\it viz.}
 \beq \text{var}(I_T^{\rm d}) \equiv (\De I_T^{\rm d})^2 = \fr{n_0 \om_0^2}{T\tau}.
 \label{DeltaId} \eeq

Thus, although the dominant phase noise component is removed when one takes the intensity difference between the beams, information about the original incident flux is {\it not}.  The minor component of photon shot noise affects both beams independently and now becomes the dominant component.  More precisely, the merit of the proposed method is that the remaining noise in the subtracted beam contains information about the photon arrival rate; from a measurement of the variance, using (\ref{DeltaId}), we can determine $n_0$.

\section{Sample variance as estimator of the population mean}

How may this approach improve the detection sensitivity of a transient or steady source?  In radio astronomy the background flux is always large, so that the phase noise is the dominant reason for the masking of faint sources and shot noise is relatively minor.  The sum of the measured $I_t$ and $I_r$ recovers the original incident intensity $I_i$, but that measurement is subject to the phase noise.  The intensity difference, on the other hand, fluctuates about zero according to the shot noise only, and the variance of this fluctuation also contains information about the signal.

The aim is to find a way of measuring as accurately as possible the incident intensity $\bar I$, or equivalently, in view of (\ref{Iav}), the value of $n_0$.  If we have available a long observation time $\cT$ then the accuracy with which we can directly measure $\bar I$ is given by (\ref{radio}).  Suppose however that instead of making this direct measurement we use the beam splitter and measure the variance of the difference signal, using a large number $N$ of short intervals $T$.  The mean of the difference signal is zero, but its variance depends linearly on $n_0$.  So the variance provides an estimator of the value of $n_0$ or equivalently the incident intensity $\bar I$.  To know how good this estimator is, we need to compute the variance of the variance.  If $N$ is large, this is given by
 \beq \text{var}[(\De I^{\rm{d}}_T)^2] \ap \fr{\mu_4 - \mu_2^2}{N}, \eeq
where $\mu_4$ is the fourth central moment of the distribution of $I^{\rm{d}}_T$, and $\mu_2$ is the second, namely $\mu_2=(\De I^{\rm{d}}_T)^2$.  If $N$ is not large, there are correction terms of order $1/N$.

The computation of the fourth moment, $\mu_4$, can be done by the same method as before, but now involves a product of eight fields rather than four.  As in the earlier calculation, the expectation value of this product factorizes into a sum of products of two-point functions.  This involves somewhat tedious algebra, but fortunately, because we are interested in the difference signal, many of these terms cancel.  The details are outlined in the appendix.  The result is
 \beq \mu_4 = \fr{n_0^2\om_0^4}{T^2\ta^2}\left[3 + 3\fr{\ta}{T} F\left(\fr{T}{\ta}\right)
 + \fr{\ta}{Tn_0} \right]. \eeq
Since $(\De I^{\rm{d}}_\cT)^2$ is proportional to $n_0$, the accuracy with which we can estimate $n_0$ by this procedure is
 \beq \fr{\De n_0}{n_0} = \fr{\De[(\De I_T^{\rm d})^2]}{(\De I_T^{\rm d})^2} =
 \fr{1}{\sqrt{N}}\left[2 + 3\fr{\ta}{T} F\left(\fr{T}{\ta}\right)
 + \fr{\ta}{Tn_0} \right]^{1/2}. \label{accdiff}\eeq

An important advantage of the difference signal is that data in non-overlapping time periods are uncorrelated, because the correlation function $\<I_T^{\rm d}(t)I_T^{\rm d}(0)\>$ is proportional to a delta function as can easily be checked by examining the integrand of the linear term in (\ref{DeltaI_T1}) which is the only one to survive the `differencing'.  So this formula should apply even if the basic time interval $T$ is very short.  The situation in which the improvement over the direct measurement is most significant is when $1/\om_0 < T <\ta$.  For small $T/\ta$ the value of $F$ is approximately $T/\ta$, so we obtain
 \beq \fr{\De n_0}{n_0} \ap \sqrt{\fr{1}{N}\left(5+\fr{\ta}{n_0 T}\right)}. \label{accsplit} \eeq
This accuracy must be compared to the value (\ref{radio}) for direct observations of the incident flux without the aid of the beam splitter.

Indeed, if a direct observation is made over the long time $NT$, then the accuracy is given by
 \beq \fr{\De n_0}{n_0} = \fr{\De I_{NT}}{\bar I} = \left(\fr{\sqrt{\pi}\ta}{NT}\right)^{1/2}.
 \label{accdir}\eeq
The ratio of the uncertainties given by the two methods, (\ref{accdiff}) and (\ref{accdir}), is
 \beq \eta = \fr{(\De n_0)_{\rm split}^2}{(\De n_0)_{\rm dir}^2}
 = \fr{1}{\sqrt{\pi}}\left[2x+3F(x)+\fr{1}{n_0}\right], \qquad x=\fr{T}{\ta}. \label{rsen} \eeq
It is a strictly increasing function of $x$, so its minimum value occurs for $x$ as small as possible. Ideally, $x=T/\ta$ should of order $1/n_0$ or smaller.  This corresponds to an interval $T$ in which of order one photon is detected, $\bar n_T \ap 1$.  Then the ratio of accuracies can be of order $1/\sqrt{n_0}$, generally a very small number for radio observations, \ie~the improvement in the signal-to-noise ratio over and above the conventional method that does not use any beam splitter is by the factor $\sqrt{n_0} \gg 1$. It can readily be verified that for all intervals $T > 1/\om_0$ the Heisenberg Uncertainty Principle $T\De E \gtrsim 1$ is satisfied.  However, because the basic approximation underlying all calculations in this paper is $T\gg 1/\om_0$ (see (\ref{IT}) and (\ref{expIsq})) it is necessary to choose a sufficiently small spectral filter bandwidth to ensure that the $\bar n_T \approx 1$ occurs {\it before}
$T$ falls below $1/\om_0$.  To elaborate, since $n_0$ is independent of the bandwidth and a narrower bandpass means larger $\ta$, one gets a smaller mean number of arriving photons in a given $T$.

As a concrete example, we take the Arecibo telescope where the system noise temperature is $\sim$ 35~K and the gain is $G \ap$ 10~K~Jy$^{-1}$  at $\nu =1$ ~GHz (1 Jansky = 10$^{-26}$~W~m$^{-2}$~Hz$^{-1}$), which converts to the flux density of 3.7~Jy , or~3.7 $\times 10^3$~photons~s$^{-1}$~Hz$^{-1}$ over the entire telescope's collecting aperture of 300~m diameter.  Assuming 50 \% detection efficiency, the occupation number of background radio photons set by (\ref{Iav}) at
 \beq n_0 = 210~{\rm at}~G = 10~{\rm K~Jy}^{-1}. \label{back} \eeq
Thus, by coupling the telescope output with the system presented here the $1\sigma$ error in one's knowledge of the background level may be reduced by as much as the factor $\sqrt{n_0} \ap 14.5$ for equivalent exposures $T$ to a source, \ie~sources fainter by as much as $14.5$ times than the detection threshold (for conventional techniques without employing the present scheme) at a given exposure may now in principle be discernible above background.

Note that because in radio astronomy the data are background dominated, $n_0$ remains $\gg 1$ irrespective of the source brightness, and thus from the analysis of the previous paragraph the improvement in signal-to-noise is always significant.  Moreover, because fast timing is the key to obtaining maximum effect, the proposed technique has application to the search for radio bursts (\cite{tho13}) of low brightness.  From (\ref{back}) and (\ref{band}) the necessary criteria for reaching the maximum signal to noise improvement factor of 14.5, {\it viz.} $\bar n_T \ap 1$ and $T \gg 1/\om_0$ (see the previous paragraph), are fulfilled by choosing a spectral filter bandpass $\de\om < \om_0/210$.

A point of clarification is in order here.  The reader should not be led to believe that the factor $\sqrt{n_0}$ means the choice of a higher equivalent noise temperature (hence $n_0$) via the use of less effective cryogenics will result in a more sensitive observation.  The reason is that in order to detect a weak source of given brightness by the conventional method (\ref{accdir}) the exposure time $NT$ has to increase with the background $n_0$ as $NT \propto n_0^2$, whereas the present method as applied to the $T \ll \ta$ regime would require the less severe $NT \propto n_0$ from (\ref{accsplit}), \ie~under either scenario the exposure time is always less for smaller $n_0$.  The other disadvantage of a high $n_0$ is that one needs faster electronics to detect down at the $\bar n_T \ap 1$ limit, and narrower bandpass to keep $T$ above $1/\om_0$; both are more technically challenging.

\section{A specific observational setup}

In terms of a realistic design, the receiver sensitivity should not be a problem to detect as few as $\sim$10 photons at 1~GHz on the timescale of 1~ns, and the pre-amplifiers and square-law detectors should be fast enough to capture the waveforms within 1~ns.
Practically, the feasibility of the proposed scheme hinges on whether one can create a `beam splitter' of radio waves that offers a stable power splitting ratio. The 180$^\circ$ hybrid junctions appear to be one of the microwave engineering techniques capable of addressing this need (\cite{poz12}). Such junctions can be fabricated into waveguide, coax or planar forms and can be inserted into the signal paths of existing radio telescopes, after the microwave preamplifiers but before the square-law detectors. Assuming the radio wave signals at this point are already converted into guided modes, such an addition should only involve minor system modifications.

For a possible working scenario, let us consider $\nu_0 =$ 1~GHz radiation from the output of the Arecibo Telescope, the details of which were presented at the end of the last section.  If a bandpass filter of $\de\nu \ap 3$~MHz is applied, and the highest timing resolution in intensity sampling is pushed right to the limit $T = 1/\om_0 \ap 0.16$~ns, the parameter $x$ in (\ref{rsen}) with $\ta$ defined in (\ref{band}) will have the minimum value of $x \ap 1/300$.  For sampling at this rate or lower, the factor of reduction in the noise {\it amplitude}, as given by $\eta^{-1/2}$ with $\eta$ defined in (\ref{rsen}), is plotted against $x$ in Figure 2.  It can be seen that at the limiting resolution the improvement over conventional sensitivity is by an order of magnitude.

\begin{figure}
\begin{center}
\includegraphics[angle=0,width=6in]{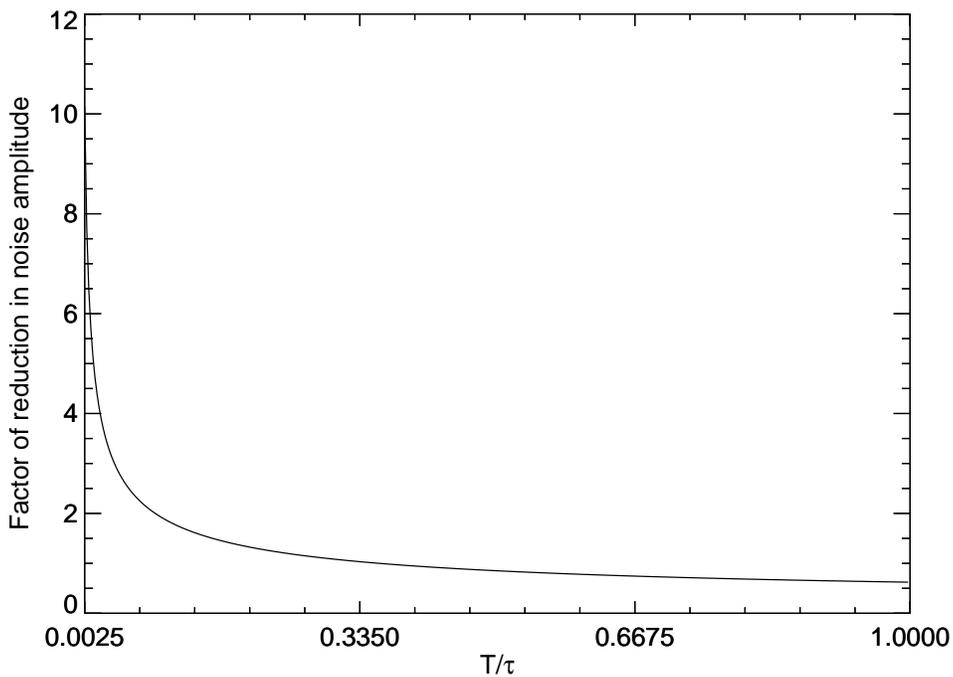}
\end{center}
\caption{The reduction factor $\eta^{-1/2}$ in noise amplitude attainable by the split beam subtraction as compared to the conventional technique is plotted against the time resolution $x$, where $\eta$ and $x$ are defined in (\ref{rsen}).  The signal is taken from the output of the Arecibo Observatory at central frequency $\nu_0 = 1$~GHz and with $n_0 \ap 210$.  The bandpass filter employed is $\de\nu/\nu_0 \ap 1/300$, and the minimum sampling resolution is $T \ap 0.16$~ns or $x=T/\ta \ap 1/300$.}
\end{figure}

\section{Conclusion}

The data gathered by radio telescopes in the frequency range 1 - 10 GHz are for the majority of sources background dominated by phase noise.  By means of a beam-splitter it is possible to remove this noise, leaving behind the much weaker fluctuations of the shot (Poisson) noise from which the incident flux can be inferred with an accuracy exceeding the conventional method of analyzing the direct beam.  The maximum improvement in sensitivity is of order $\sqrt{n_0}$ where $n_0$ is the photon occupation number, and is attained when the time resolution is finer than the coherence time, reaching the limit of one arriving photon per resolution element on average. Since radio frequencies are low, the fast timing devices necessary to put this idea into action are relatively easy to build.

\appendix

\section{Computation of fourth moment}

To compute the variance of the variance of $I_T^{\rm d}$ we require its fourth central moment $\mu_4$.  Now
 \bea \hat I_T^{\rm d} \!\!\!&=&\!\!\! \fr{2A}{T} \int_0^T dt_1\,
 [\hat E^{{\rm t}(-)}(t_1)\hat E^{{\rm t}(+)}(t_1)-\hat E^{{\rm r}(-)}(t_1)\hat E^{{\rm r}(+)}(t_1)] \notag \\
 \!\!\!&=&\!\!\! \fr{1}{2\pi T} \int_0^T dt_1\, \int d\om_1\,d\om'_1\sqrt{\om_1\om'_1}e^{i(\om_1-\om'_1)t}
 [\hat a^{{\rm t}\dag}(\om_1)\hat a^{{\rm t}}(\om'_1)
 -\hat a^{{\rm r}\dag}(\om_1)\hat a^{{\rm r}}(\om'_1)]. \eea
We need to compute the expectation value of the product of four these operators, for which we will replace the subscript 1 by 2,3,4.  This involves the expectation value of a product of eight $\hat a$ and $\hat a^\dag$ operators, which we may denote in a self-explanatory notation by the symbol
$\<11'22'33'44'\>$.  As before, for a Gaussian field, this expectation value factorizes into a sum of products of two-point functions, e.g., $\<11'\>\<23'\>\<34'\>\<2'4\>$.  There are 24 terms in this sum, corresponding to the different ways of pairing the four $\hat a^\dag$ operators with the four $\hat a$ operators.  Each can be associated with the corresponding permutation of the primed symbols, e.g., in this case, using the cycle notation the permutation $(1)(234)$.  Each of the 24 terms itself consists of a sum of 16 terms, representing the t or r choices for each pair of fields.  Fortunately, as we shall see, many of these terms cancel out.

Firstly, any term containing a pairing such as $\<11'\>$ will vanish because the t and r terms cancel, as in the expectation value of $\hat I^{\rm d}$.  So we can drop all permutations containing (1) for example.  Thus we only need to consider permutations belonging to the partitions $[2^2]$ or $[4]$.  Any of the three $[2^2]$ permutations, for example $(13)(24)$, can easily be seen to give a result that is simply $\<(\hat I_T^{\rm d})^2\>^2$.  So we are left with permutations of class $[4]$.

Next, for any pairing that is in normal order, such as $\<12'\>$, this factor has the same magnitude irrespective of t or r labels, namely $\half n(\om_1)\de(\om_1-\om'_2)$, but terms that differ only by replacing one label t by r or vice versa appear with opposite signs.  Thus, if for any of the digits, both the relevant pairs are in normal order, for example for the digit 2, $\<12'\>$ and $\<24'\>$, then it is clear that any two terms in the 16 that differ by replacing t by r for the $2,2'$ factors will be identical except for sign, and therefore cancel.  It is not difficult to see that among the permutations of class $[4]$, these conditions eliminate all but three, namely $(1324),(1423)$ and $(1432)$.

For pairings that are not in normal order, such as $\<1'2\>$, the situation is more complicated.  We have
 \beq \<1'2\>^{\rm{tt}}=\<1'2\>^{\rm{rr}}=[\half n(\om_1)+1]\de(\om'_1-\om_2), \eeq
but
 \beq \<1'2\>^{\rm{tr}}=\<1'2\>^{\rm{rt}}=\half n(\om_1)\de(\om'_1-\om_2). \eeq
Thus the cancellations are incomplete.  For (1324), where two of the factors are in normal order and two in antinormal order we obtain for the sum of all 16 choices of t and r the expression
 \beq \<13'\>\<2'3\>\<24'\>\<1'4\> = n(\om_1)n(\om_2)
 \de(\om_1-\om'_3)\de(\om'_2-\om_3)\de(\om_2-\om'_4)\de(\om'_1-\om_4). \eeq
In the contribution of this term to the integral representing $\mu_4$ we can use the delta functions to perform the $\om'_j$ integrations and obtain
 \bea &&\fr{1}{(2\pi)^4T^4} \int_0^T dt_1dt_2dt_3dt_4 \int d\om_1 d\om_2 d\om_3 d\om_4\notag\\
 &&\qquad \om_1\om_2\om_3\om_4 n(\om_1)n(\om_2)
 e^{i[(\om_1-\om_4)t_1+(\om_2-\om_3)t_2+(\om_3-\om_1)t_3+i(\om_4-\om_2)t_4]}. \eea
As before, the $\om_j$ factors can all be replaced to a very good approximation by $\om_0$.  In the $\om_3$ and $\om_4$ integrations, integrating by parts, these yield extra factors of $\om_j$ or equivalently $\om_0$, together with delta functions $\de(t_3-t_2)\de(t_4-t_1)$.  Thus we obtain
 \bea && \fr{\om_0^4}{(2\pi)^2T^4} \int_0^T dt_1dt_2 \int d\om_1 d\om_2\
 n(\om_1)n(\om_2) e^{i(\om_1-\om_2)(t_1-t_2)} \notag\\
 && \qquad = \fr{1}{(2\pi)^2 T^4} \int_{-T}^T dt\,
 (T-|t|)\left|\int d\om\,\om^2 n(\om)e^{i\om t}\right|^2 \notag\\
 && \qquad = \fr{\om_0^4n_0^2}{T^4\ta^2}2\int_0^T dt\,(T-t)e^{-t^2/\ta^2}
 = \fr{\om_0^4n_0^2}{T^3\ta}F\left(\fr{T}{\ta}\right), \eea
where $F(x)$ is the same function defined in (\ref{F}).

For (1423), the calculation is exactly the same apart from relabelling of variables, and the answer is also the same.

Finally we are left with (1432), where \emph{three} of the pairs are in antinormal order.  In this case, it turns out that we have not only the same terms involving two $n$ factors, but also one involving just one.  We get for the sum of the sixteen terms with all possible choices of t and r,
 \beq \<14'\>\<3'4\>\<2'3\>\<1'2\> = n(\om_1)[n(\om_3)+1]
 \de(\om_1-\om'_4)\de(\om'_3-\om_4)\de(\om'_2-\om_3)\de(\om'_1-\om_2). \eeq
The term proportional to $n(\om_3)$ contributes exactly the same as for the two preceding terms, but we now have an extra term arising from the $(+1)$.  This is very similar to the extra shot-noise  term that appeared in (\ref{DeltaI_T}).  The net contribution to $\mu_4$ is
 \beq \fr{\om_0^4}{T^3\ta}\left[n_0^2 F\left(\fr{T}{\ta}\right)+n_0\right]. \eeq

Finally, if we put all these terms together, we obtain
 \beq \mu_4 = \fr{n_0^2\om_0^4}{T^2\ta^2}\left[3 + 3\fr{\ta}{T} F\left(\fr{T}{\ta}\right)
 + \fr{\ta}{Tn_0} \right]. \eeq


\begin{thebibliography}{}

\bibitem[Burke \& Graham-Smith (2010)]{bur10} Burke, B.F., \& Graham-Smith, F., 2010, An Introduction to Radio Astronomy, 3rd edition, Cambridge University Press.

\bibitem[Birney et al (2006)]{bir06} Birney, D.S., Gonzalez, G., \& Oesper, D., 2006, Observational Astronomy, 2nd edition, Cambridge University Press.

\bibitem[Christiansen \& H\"ogbom (1985)]{chr85} Christiansen, W.N., \& H\"ogbom, J.A., 1985, Radio Telescopes, 2nd edition, Cambridge University Press.

\bibitem[Hanbury-Brown \& Twiss (1957)]{han57} Hanbury Brown, R., \& Twiss, R.Q., 1957, Proc. Roy. Soc. A, 242, 300

\bibitem[Pozar (2012)]{poz12} Pozar, D. M., 2012, Microwave Engineering, 4th edition, John Wiley \& Sons.

\bibitem[Thornton et al (2013)]{tho13} Thornton, D., et al, 2013, Science, 341, 53

\end{thebibliography}
\end{document}